\documentclass[preprint,aps]{revtex4}
\usepackage{mathrsfs}
\usepackage{graphicx}

\begin{document}
\title{Preparation of superconducting thin films of infinite-layer nickelate Nd$ _{0.8}$Sr$ _{0.2}$NiO$ _{2}$ }

\author{Qiang Gao$^{1}$, Yuchen Zhao$^{1,2}$, Xingjiang Zhou$^{1,2,3,4,*}$, and Zhihai Zhu$^{1,2,*}$}

\affiliation{
\\$^{1}$National Lab for Superconductivity, Beijing National laboratory for Condensed Matter Physics, Institute of Physics,
Chinese Academy of Sciences, Beijing 100190, China
\\$^{2}$School of Physical Sciences, University of Chinese Academy of Sciences, Beijing 100049, China
\\$^{3}$Beijing Academy of Quantum Information Sciences, Beijing 100193, China
\\$^{4}$Songshan Lake Materials Laboratory, Dongguan 523808, China
\\$^{*}$Corresponding author:  XJZhou@iphy.ac.cn, zzh@iphy.ac.cn
}

\date{\today}





\maketitle

\newpage

{\bf The recent observation of superconductivity in thin films of infinite-layer nickelate Nd$ _{0.8}$Sr$ _{0.2}$NiO$ _{2}$ has received considerable attention. Despite the many efforts to understand the superconductivity in infinite-layer nickelates, a consensus on the underlying mechanism for the superconductivity has yet to be reached, partly owing to the challenges with the material synthesis. Here, we report the successful growth of superconducting infinite-layer Nd$ _{0.8}$Sr$ _{0.2}$NiO$ _{2}$ films by pulsed laser deposition and soft chemical reduction. The details on growth process will be discussed. }


Despite decades of research on cuprate superconductors, an understanding of what exact features of a material essentially support superconductivity has remained elusive.  On an empirical basis, it seems natural to expect that superconductivity can be achieved through synthesis of cuprate analogs that share common features including spin one-half, two-dimensionality, and strong antiferromagnetic correlations etc \cite{GKhaliullin2008JChaloupka}. Along this direction, Chaloupka et al. proposed that LaNiO$_3$/LaMO$_3$ (M stands for a trivalent cation Al, Ga, etc) superlattices would create superconductivity, which however has not been realized to date although the nickel-oxide superlattices have been successfully synthesized \cite{BKeimer2011AVBoris}. The recent report of superconductivity in an infinite-layer nickelate Nd$ _{0.8}$Sr$ _{0.2}$NiO$ _{2}$ has opened an alternative route to achieve superconductivity in transition metal oxide materials in addition to cuprates \cite{HYHwang2019DFLi}. Compared with cuprates, the superconducting nickelates comprise a similar two-dimensional (2D) square lattice, isoelectronic 3\textit{d}$^9$ valence state, and a superconducting dome in phase diagram \cite{HYHwang2020DFLi,AAriando2020SWZeng,HYHwang2020MOsadaPRM,WSLee2020MHepting,LFKourkoutis2021BHGoodge,YDChuang2020MRossi}. But in sharp contrast to cuprates, the parent compound NdNiO$ _{2}$ of superconducting nickelates does not exhibit any magnetic order down to 1.7 K \cite{MJRosseinsky2003MAHayward}, indicating the underlying superconducting mechanism differs from that of cuprate superconductors for which magnetism are generally believed to be crucial for mediating Cooper pairs. Although various theoretical models have been proposed to explain the superconductivity in the superconducting nickelates \cite{MRNorman2020ASBotana,FMarsiglio2019JEHirsch,ZCZhong2019PHJiang,RArita2019YNomura,
FCZhang2020GMZhang,TSaha2020PAdhikary,AMillis2020JKarp,TSaha2020SBandyopadhyay,YDing2020JChang,
WEPickett2020MYChoi,RPentcheva2020BGeisler,HHChen2020YHGuBulletin,HHChen2020YHGuComm,ZCZhong2020RHe,
XPYang2020CJi,GASawatzky2020MJiang,ASBotana2020JKapeghian,AAlavi2020VMKatukuri,
ASBotana2020JKrishna,WKu2020ZJLang,HMWeng2019JCGao,FLechermann2020PRX,FLechermann2020PRB,SYSavrasov2020ILeonov,JLYang2020ZLIu,
OErten2020EMNica,PWerner2020FPetocchi,MJHan2020SRyee,KKuroki2020HSakakivara,KHeld2020LSi,GKotliar2020YWang,
FCZhang2020ZWang,SHoshino2020Werner,RThomale2020XXWuPRB,RThomale2020XXWuarxiv,JWSun2020RQZhang,EDagotto2020YZhang,SRachel2021MKlett,
JLYang2021ZLiu,SYSavrasov2021XGWan}, a consensus model has not yet reached partly owing to the challenges with the synthesis, and only very few reports exist on achieving superconducting films of infinite-layer nickelates \cite{HYHwang2019DFLi,AAriando2020SWZeng,HYHwang2020MOsada,HYHwang2020KLee,HHWen2020QQGu,HHWen2020YXiang,HYHwang2021BYWang,ZQLiu2020XRZhou}.  At present, superconductivity has not been observed in bulk Nd$_{1-x}$Sr$_x$NiO$_2$ \cite{HHWen2020QLI,DPhelan2020BXWang,HHWen2020CPHe,WQYu2020YCui}. 

In this paper, we report the successful growth of superconducting infinite-layer Nd$ _{0.8}$Sr$ _{0.2}$NiO$ _{2}$ films. The perovskite precursor phase nickelate films were prepared by using pulsed laser deposition (PLD). The infinite-layer phase was acquired by soft-chemistry reduction method.  The thickness and out-plane x-ray diffraction (XRD) pattern of the prepared films  were examined using a SmartLab x-ray diffractometer. The superconductivity was confirmed by transport measurements using a Quantum Design Physical Property Measurement System (PPMS) with a standard four-probe configuration. The wire connection was made by melting indium.

The laser target contained a stoichiometric mixture of SrCO$_3$ (Alfa Aesar, 99.99\%), Nd$_2$O$_3$ (Alfa Aesar, 99.99\%),  and NiO (Sigma Aldrich-Chemie GmbH, 99.995\%) prepared by a solid-state reaction in the air at 1100 $^{\circ}$C for 24 hours. The products of this reaction were ground and reheated, and this process was repeated five times.  The resulting polycrystalline materials was pressed into a pellet and sintered for 24 hours at 1200 $^{\circ}$C in the air. The heating and cooling rate of sintering were kept at 3 $^{\circ}$C/min. 

The nickelate films were grown by PLD using 248-nm KrF excimer laser (COMPex 201, Coherent). The SrTiO$ _3$ (001) substrates (5*5 mm, without chemical etching to achieve a TiO$ _2$ terminated surface) were pre-annealed at 900 $^{\circ}$C with an oxygen partial pressure of $1 \times 10^{-5}$ Torr. During growth, the substrate temperature was kept at 600  $^{\circ}$C under an oxygen partial pressure of 150 mTorr. After deposition, the films were cooled to room temperature at a rate of 5 $^{\circ}$C per minute in the same oxygen partial pressure. The laser beam size was about $0.8 \times 3.2$ mm$^2$ realized by using an aperture. The pulse energy of the laser was set to 430 mJ  and 730 mJ for the growth of NdNiO$ _{3}$ and Nd$ _{0.8}$Sr$ _{0.2}$NiO$ _{3}$,  respectively. The laser frequency was set to 4 Hz.

The infinite-layer nickelate phase was acquired by soft-chemistry reduction method.  As shown in Fig. 1 (b), the as-grown nickelate films were wrapped in clean aluminum foil and then sealed with 0.1 g CaH$ _2$ powder (Alfa Aesar, 98{\%}) in quartz tubes which were pumped to a vacuum better than $1 \times 10^{-5}$ Torr. The reduction was carried out at a temperature of 290 $^{\circ}$C for 5 hours, with the heating and cooling rates of 10 $^{\circ}$C/min.

Figure \ref{Fig2} shows the structural characterization of the nickelate films. The NdNiO$_3$ film peaks at 23.5$^{\circ}$ and 48.2$^{\circ}$ correspond to the (001) and (002) reflections, respectively (see Fig.2 (a)). After chemical reduction, as shown in Fig. 2 (b) the film peaks at 26.6$^{\circ}$ and 55.7$^{\circ}$ identify the infinite-layer NdNiO$_2$ and correspond to (001) and (002) reflections, respectively.  For a typical film of Nd$_{0.8}$Sr$_{0.2}$NiO$_{3}$ shown in Fig. 2(c), the  peaks at 23.8$^{\circ}$ and  48.4$^{\circ}$ correspond to (001) and (002) reflections, respectively, and both two peaks are slightly broader than those of undoped phase in Fig. 2 (a).  After chemical reduction, the peaks at 26.5$^{\circ}$ and 54.8$^{\circ}$ correspond to (001) and (002) reflections, respectively, as expected for a film of an infinite-layer Nd$_{0.8}$Sr$_{0.2}$NiO$_{2}$ (Fig.\ref{Fig2} (d)). The intensity is comparable to the precursor. It was reported that the Nd$ _{0.8}$Sr$ _{0.2}$NiO$ _{3}$ films were difficult to grow due to the formation of a secondary phase which shows only a single peak with 2$\theta$ less than 48$^{\circ}$ \cite{HYHwang2020KLee}.

In Fig.\ref{Fig3} we show the measured Nd$ _{0.8}$Sr$ _{0.2}$NiO$ _{2}$ film resistance as a function of temperature revealing that  a superconducting transition occurs below 9 K and the zero resistivity is achieved at about 3.5 K. The superconducting transition temperature of our film is a little lower than that reported in the literature,  which probably stems from the variation of the hole concentration in the films prepared under differing growth and reduction conditions. To study the stability of the superconducting Nd$_{0.8}$Sr$_{0.2}$NiO$_2$ films, we measured the resistivity of a typical film which had been kept in the glovebox for 30 days after chemical reduction and had been exposed to air longer than 10 hours during XRD and other measurements. The superconducting transition is almost identical with the same critical temperature $T_c$, as shown in Fig.\ref{Fig3}(c).  It seems that the 8.5nm-thick superconducting film without capping layer is fairly stable, promising the suitability for multiple experiments that require stable samples. 


Currently, using the PLD growth and soft chemical reduction procedure introduced above, we manage to produce at least one superconducting sample out of the ten films, and we are continuing our optimization of the growth condition to improve even higher the reproducibility. The growth window is extremely narrow probably because the formation energy of the desired 113 phase differs only slightly from that of the secondary phase. Nevertheless, we find that the XRD pattern of the precursor 113 phase is a good criterion to obtain superconducting films. The (002) peak of the secondary phase locates at about 47.8$^{\circ}$, while for 113 phase the (002) peak is at the 2$\theta$ larger than 48$^{\circ}$ and thickness dependent. In the optimization, the 2$\theta$ of the  (002) reflection varies almost continuously from 47.8$^{\circ}$ to 48.5$^{\circ}$. 
For a typical 8.5nm-thick  Nd$ _{0.8}$Sr$ _{0.2}$NiO$ _{3}$ film, we find that as long as the (002) peak reaches 48.5$^{\circ}$, the superconducting phase is easily achieved through chemical reduction; and if the (002) peak locates at 48.3$^{\circ}$, the superconductivity can be achieved, though it is much harder to be turned into the infinite-layer phase using the chemical reduction, and the corresponding critical temperature Tc varies from 5K to 9K. However, the precursor films with 2$\theta$ of the (002) peak less than 48$^{\circ}$ have never been reduced to the infinite-layer phase,  let alone the superconducting phase.


In summary, we have successfully synthesized superconducting infinite-layer Nd$ _{0.8}$Sr$ _{0.2}$NiO$ _{2}$ thin films by using pulsed laser deposition and soft-chemistry reduction method. The details on the film growth and subsequent chemical reduction process are discussed. Our results provide important information for the preparation of superconducting nickelate films that is a crucial step in this field.

\vspace{3mm}

\noindent {\bf Acknowledgement}\\
We wish to thank Er-jia Guo for helpful discussions. This work was supported in part by the National Natural Science Foundation of China (Grant No. 12074411) and  (Grant No. 11888101), the National Key Research and Development Program of China (Grant No. 2016YFA0300300 and 2017YFA0302900), the Strategic Priority Research Program (B) of the Chinese Academy of Sciences (Grant No. XDB25000000) and the Research Program of Beijing Academy of Quantum Information Sciences (Grant No. Y18G06).

\vspace{3mm}

\noindent {\bf Author Contributions}\\
 Z.H.Z., X.J.Z., and Q.G. proposed and designed the research. Q.G., Y.C.Z. and Z.H.Z. grew the films. Q.G. carried out the XRD and transport measurements with the help from Y.C.Z. Z.H.Z. and Q.G. wrote the paper. All authors participated in discussions and comments on the paper.

\vspace{3mm}


\vspace{3mm}

\bibliographystyle{unsrt}
\bibliography{ref}

\newpage

\begin{figure*}[tbp]
\begin{center}
\includegraphics[width=1.0\columnwidth,angle=0]{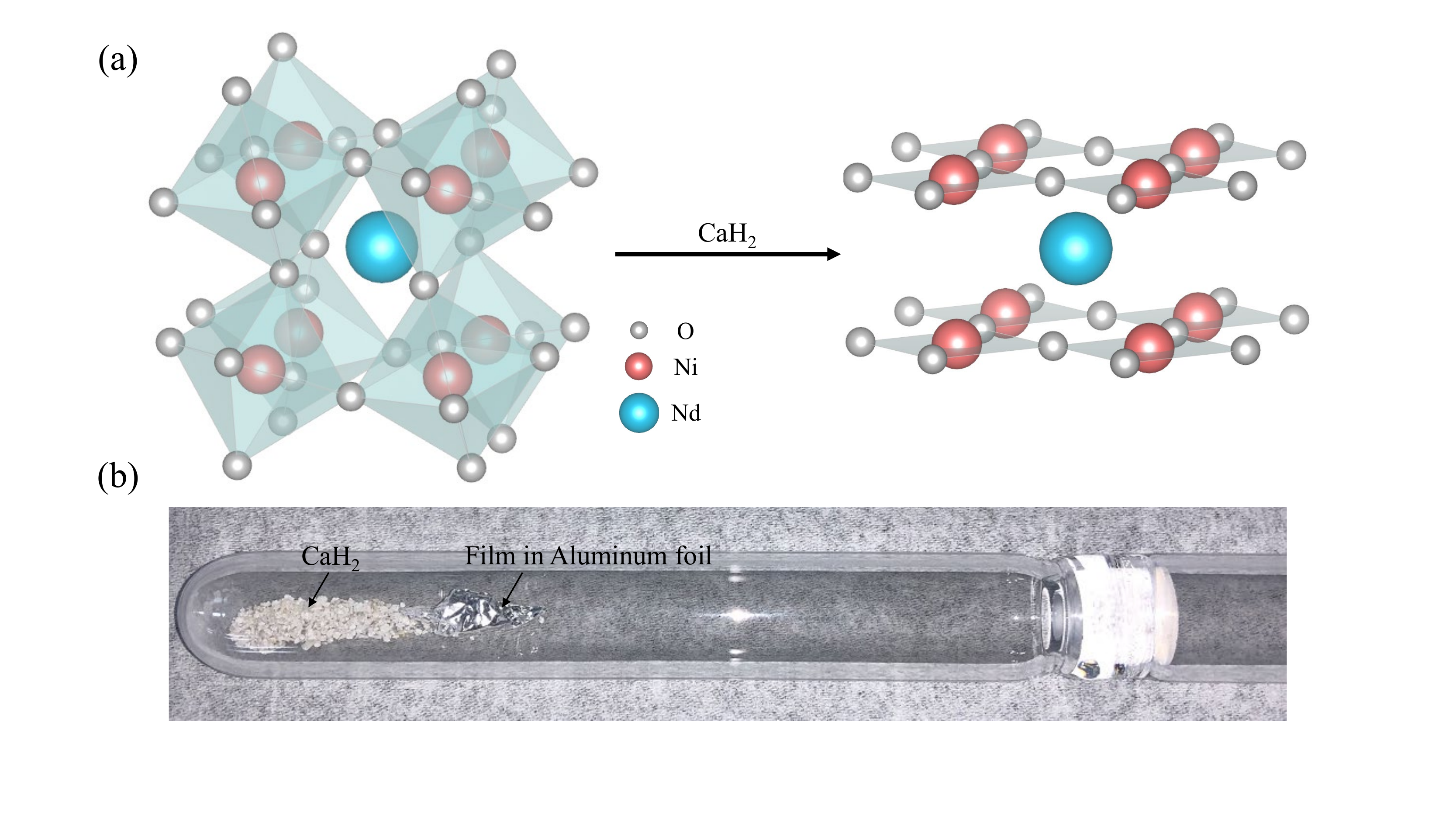}
\end{center}
\caption{{\bf Crystal structure and the reduction process of nickelates. } (a) illustrates  the crystal structures of NdNiO$ _{3}$ (left) and NdNiO$ _{2}$ (right). (b) The soft-chemistry reduction process. A typical  Nd$ _{0.8}$Sr$ _{0.2}$NiO$ _{3}$ film wrapped in Aluminum foil is vacuum-sealed with CaH$_{2}$ powder in a quartz tube.}
\label{Fig1}
\end{figure*}

\begin{figure*}[tbp]
\begin{center}
\includegraphics[width=1.0\columnwidth,angle=0]{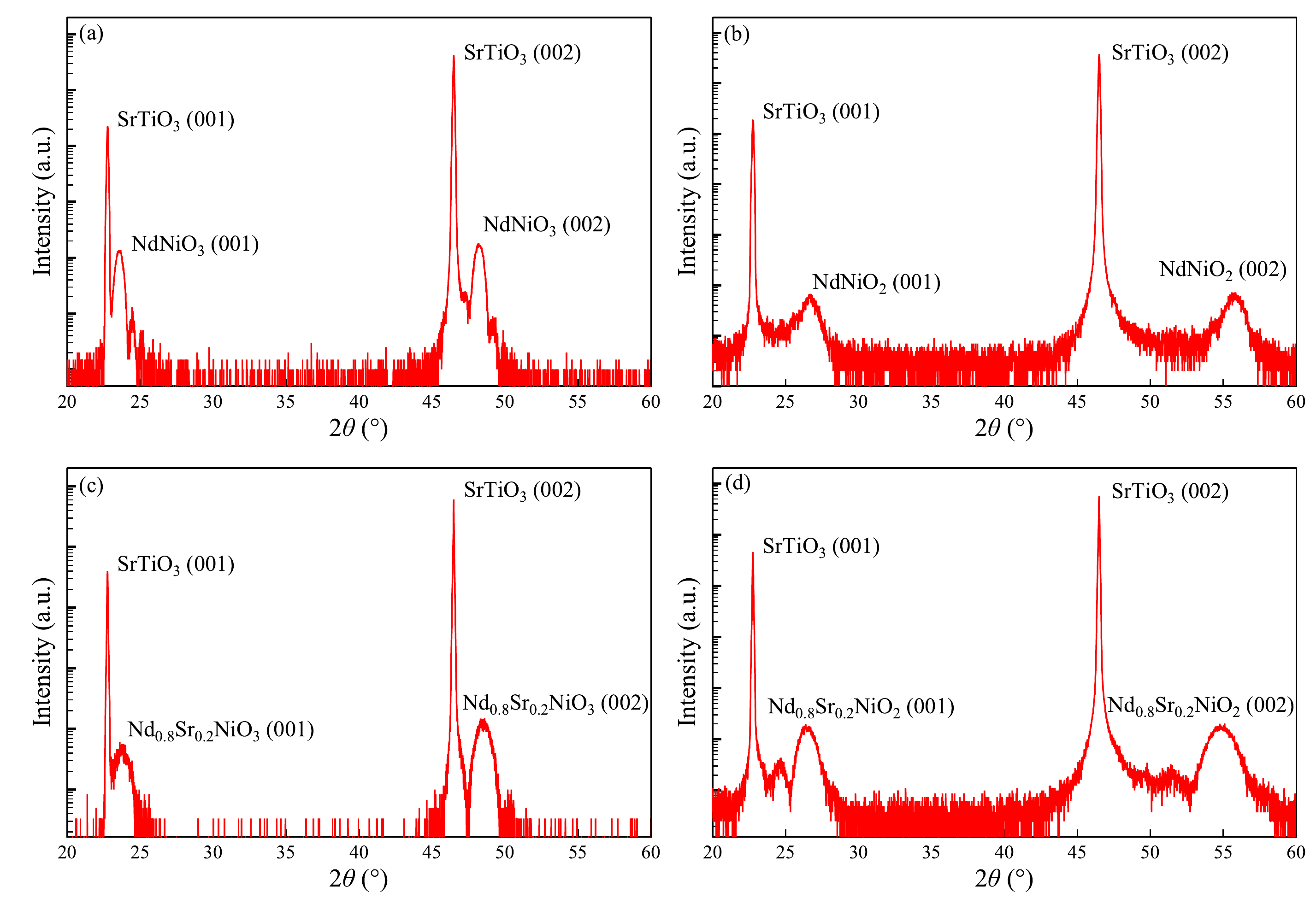}
\end{center}
\caption{{\bf Structural characterization of nickelate films on STO substrates. } (a) The XRD $\theta-2\theta$ scans of a typical NdNiO$ _{3}$ film with a thickness of 14 nm. (b) The XRD $\theta-2\theta$ scans of the infinite-layer NdNiO$ _{2}$
 film acquired by performing chemical reduction on the sample (a) at 290 $^{\circ}$C for 5 hours. (c) The XRD $\theta-2\theta$ scans of a typical Nd$ _{0.8}$Sr$ _{0.2}$NiO$ _{3}$ film with a thickness of 9 nm. (d) The XRD $\theta-2\theta$ scans of the infinite-layer Nd$ _{0.8}$Sr$ _{0.2}$NiO$ _{2}$ film acquired by reducing the sample (c) at 290 $^{\circ}$C for 5 hours.}
\label{Fig2}
\end{figure*}

\begin{figure*}[tbp]
\begin{center}
\includegraphics[width=1\columnwidth,angle=0]{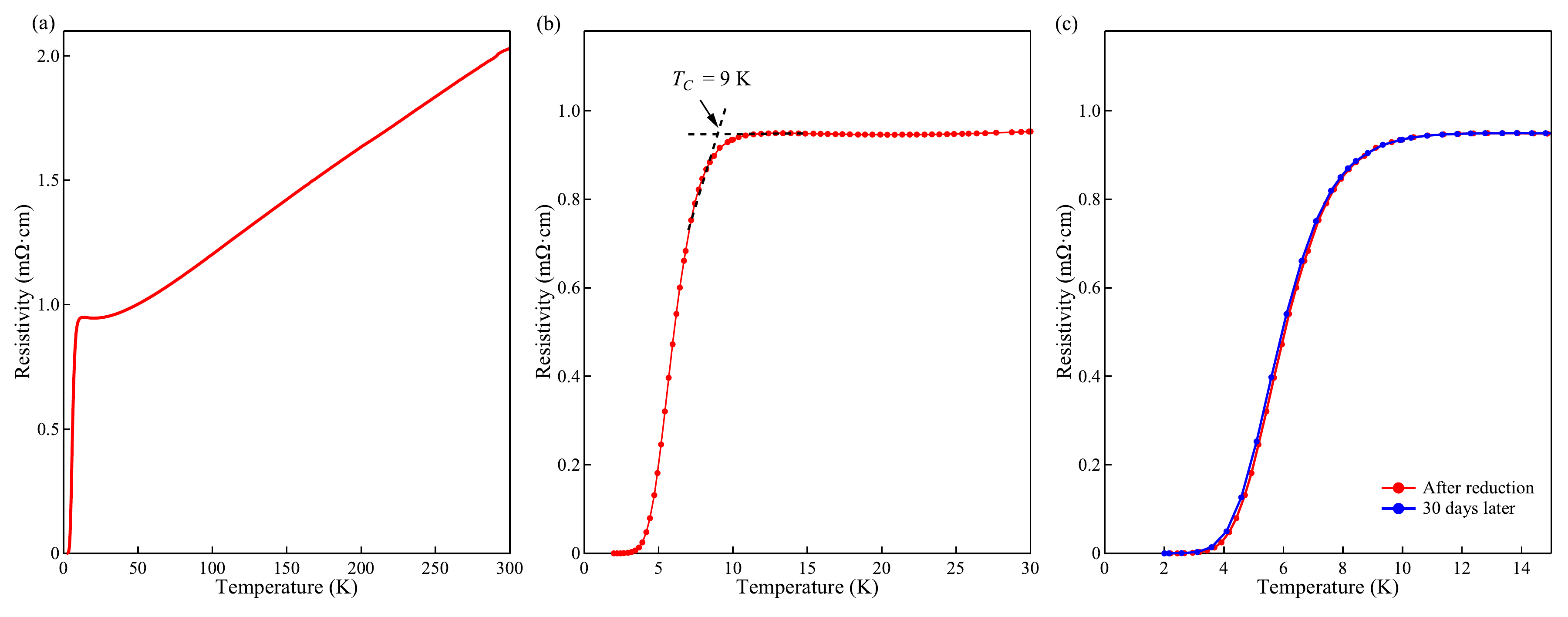}
\end{center}
\caption{{\bf Transport properties of Nd$ _{0.8}$Sr$ _{0.2}$NiO$ _{2}$ thin film.} (a) Temperature-dependent resistivity of the Nd$ _{0.8}$Sr$ _{0.2}$NiO$ _{2}$ thin film from 300 K  to 2 K. (b) A zoom-in view of (a) with the temperature lower than 30 K. The onset of the superconducting transition is about 9 K. (c) An aging test of a nickelate superconducting film. The transport measurement shows almost identical superconducting transition, over a period of 30 days after the chemical reduction.}
\label{Fig3}
\end{figure*}

\end{document}